\documentclass[twocolumn,twoside,slac_two]{revtex4}
\usepackage{graphicx}
\usepackage{fancyhdr}
\usepackage{longtable}
\newcommand{\be}{\beta}

\newcommand{\ep}{\epsilon}

\def\be{\begin{equation}}
\def\ee{\end{equation}}

\begin{document}

\title{Stability of strange stars (SS) under radial oscillation.}

%

\author{Monika Sinha}
\affiliation{Dept. of Physics, Presidency College, 86/1 College
Street, Kolkata 700 073, India, CSIR-NET fellow, Govt. of India.}
\author{Jishnu Dey}
\affiliation{UGC Research Professor, Dept. of Physics, Maulana
Azad College, 8 Rafi Ahmed Kidwai Road, Kolkata 700 013, India.}
\author{Mira Dey}
\affiliation{Dept. of Physics, Presidency College, 86/1 College
Street, Kolkata 700 073, India.}
\author{Subharthi Ray}
\affiliation{Inter University Centre for Astronomy and
Astrophysics, Post Bag 4, Ganeshkhind, Pune 411007, India.}
\author{Siddhartha Bhowmick}
\affiliation{Department of Physics, Barasat Govt. College,
Barasat, North 24 Parganas, W. Bengal, India.}

\begin{abstract}
A realistic Equation of State (EOS) leads to strange stars (ReSS)
which are compact in the mass radius plot, close to the
Schwarzchild limiting line \cite{d98}. We carry out a stability
analysis under radial oscillations and compare with the EOS of
other SS models. We find that the ReSS is stable and an M-R
region can be identified to that effect.

\end{abstract}

\maketitle

\thispagestyle{fancy}


\section{Introduction}
The radial mode oscillation, being the simplest mode of neutron
star has been considered first to be investigated more than 35
years ago \cite{chan}. It can give information about stability of
the stellar object under consideration.

The radial modes of neutron star have been studied thoroughly by
many authors for cold nuclear matter EOS
\cite{harri,chanm,glass,vath,kokkotas}. With this radial modes
have been investigated for other type of star, namely Strange Star
\cite{ben} protoneutron star \cite{gon}, Hybrid Star \cite{gupta}.

We here present our analysis of radial mode oscillation for
Realistic Strange Star (ReSS) Equation of State (EOS).

\section{ Radial oscillations of a relativistic star}

Thirty five years ago Chandrasekhar \cite{chan} investigated these
radial modes. Following him we investigate the same for ReSS.

The spherically symmetric metric is given by the line element \be
ds^2 = -e^{2 \nu } dt^2 + e^{2 \mu} dr^2 + r^2 (d \theta^2 +
\sin^2 \theta d\phi^2) . \ee

Together with the energy-momentum tensor for a perfect fluid,
Einstein's field equations yield the Tolman-Oppenheimer-Volkoff
(TOV) equations which can be solved if we have an EOS, $p(n_B)$
and $\ep(n_B)$. Given the central density $\ep_c$, we can arrive
at an $M-R$ curve by solving the TOV. Without disturbing the
spherical symmetry of the background we define $\delta r(r,t)$, a
time dependent radial displacement of a fluid element located at
the position $r$ in the unperturbed model which assumes a
harmonic time dependence, as

\be \delta r(r,t)= u_{n}(r) e^{i\omega_{n} t}. \ee

    The dynamical equation governing the stellar pulsation in its
$n$th normal mode ($n=0$, is the fundamental mode) has the
Sturm-Liouville's form (for details, see \cite{mis}).

\be P(r) \frac{d^2 u_{n}(r)}{dr^2} + \frac{dP}{dr}
\frac{du_{n}}{dr} + \left[Q(r) + \omega_{n}^2 W(r)\right] u_{n}
(r) = 0, \ee where $u_{n}(r)$ and $\omega_{n}$ are the amplitude
and frequency of the $n$th normal mode, respectively. The
functions $P(r)$, $Q(r)$ and $W(r)$ are expressed in terms of the
equilibrium configuration of the star and are given by

\be P(r) = \frac{\Gamma p}{r^2}  e^{\mu+3\nu} \ee

\be Q(r) = e^{\mu+3\nu}\left[\frac{(p')^2}{r^2(\ep+p)} -
\frac{4p'}{r^3} - \frac{8\pi}{r^2} (\ep + p) ~ p ~e^{2\mu}\right]
\ee

\be W = \frac{(\ep+p)}{r^2}~ e^{3\mu +\nu}, \ee where the varying
adiabatic index $\Gamma$ is  given by

\be \Gamma = \frac{(\ep+p)}{p}\frac{dp}{d\ep}, \ee $\ep$ and $p$
being the energy density and pressure of the unperturbed model,
respectively. Eigenfrequencies can be obtained with the boundary
conditions,
\begin{enumerate}
\item at the centre $r=0$, $\delta r= 0$ and
\item at the surface $\delta p =0$ leading to $\Gamma ~p~
u(r)'~=~0$,
\end{enumerate}

\begin{figure}[h]
\centering
\includegraphics[width=8cm]{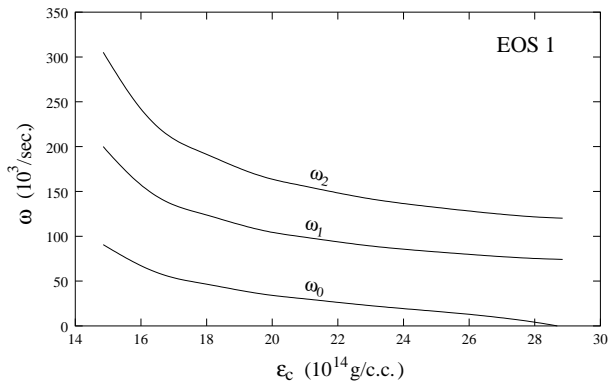}
\caption{Angular frequency of three different modes against
central density for SS1.} \label{freq}
\end{figure}

Since $\omega $ is real for $\omega ^2 > 0$, the solution is
oscillatory. However for $\omega ^2 < 0$, the angular frequency
$\omega$ is imaginary, which corresponds to an exponentially
growing solution. This means that for negative values of $\omega
^2 $ the radial oscillations are unstable. For a compact star the
fundamental mode $\omega_0$ becomes imaginary at some central
density $\ep_c$ less than the critical density $\ep_{critical}$
for which the total mass $M$ is a maximum. At $\ep_c=\ep_c^0$,
$\omega_0$ vanishes. All higher modes are zero at even higher
central densities. Therefore,  the star is unstable for central
densities greater than $\ep_c^0$. To illustrate, we plot the
eigen frequencies  $\omega_n$ against $\ep_c$, the central density
in Fig. \ref{freq}. The fundamental frequency $\omega_0$ does
vanish at some $\ep_c^0$ while the higher modes remain nonzero.

Numerical values of  masses, radii, central densities and the
corresponding eigen frequencies $\omega_0$, $\omega_1 $ and
$\omega_2$ are given in Tables \ref{eos1} and \ref{eos3} for EOS1
and EOS3 respectively (SS1 and SS2 of Dey et al. 1998). Tables
\ref{bag1} and \ref{bag2} are for the bag model EOS with different
parameters.

\section{Discussions and summary}

ReSS are stable against radial oscillations close to the maximum
attainable mass. For example, the EOS of SS1 sustains
gravitationally, $M_{max} \sim 1.4 M_{\odot}$, R=7 km with a
central number density $n_c ~\sim 16 n_0$. However, the
fundamental frequency of radial oscillations becomes zero at
around $n_c~9.5\sim n_0$,  destabilizing the star after M=1.36
$M_{\odot}$ with R= 7.24 km  (Table \ref{eos1}). It is still on
the $\frac{dM}{dR}>0$ region. Thus the maximum mass star which is
stable against radial oscillations has a  number density $\sim 9.5
n_0$ at the centre and $\sim 4.7 n_0$ at the surface.
Macroscopically, upto this density small vibrations may be
sustained.

\begin{table*}[h]
\begin{center}
\caption{Data for EOS1 (SS1)}
\begin{tabular}{|c|c|c|c|c|c|c|}
\hline
$\rho_c~10^{14}$&$n_c/n_0$&$M/M_{\odot}$&R&$\omega_0~10^3$&$\omega_1~10^3$&$\omega_2~10^3$\\
g/c.c.&&&km&/sec.&/sec.&/sec.\\
\hline
14.85&5.462&0.407&5.262&90.661&199.957&305.202\\
15.85&5.798&0.502&5.611&79.884&179.851&275.679\\
16.85&6.122&0.787&5.940&69.870&162.018&249.011\\
17.85&6.429&0.893&6.643&47.528&125.267&193.718\\
18.85&6.735&0.991&6.828&40.784&114.643&178.073\\
19.85&7.036&1.077&6.970&34.751&105.455&165.102\\
20.85&7.321&1.133&7.050&30.694&99.662&156.925\\
21.85&7.605&1.182&7.113&26.848&94.509&149.451\\
22.85&7.886&1.226&7.161&23.077&89.825&142.620\\
23.85&8.159&1.261&7.193&19.749&86.070&137.200\\
24.85&8.427&1.288&7.214&16.693&83.005&132.804\\
25.85&8.692&1.312&7.228&13.435&80.192&128.753\\
26.85&8.955&1.333&7.236&9.648&77.588&135.005\\
27.85&9.212&1.349&7.240&4.943&75.483&122.982\\
28.85&9.466&1.363&7.240&$5.899$&73.592&119.295\\
30.85&9.969&1.381&7.235&$-$&70.168&114.425\\
35.85&11.176&1.417&7.194&$-$&64.144&105.945\\
40.85&12.333&1.433&7.130&$-$&59.935&100.169\\
46.85&13.669&1.437&7.055&$-$&56.349&95.361\\
\hline
\end{tabular}
\label{eos1}
\end{center}
\end{table*}

\begin{table*}[h]
\begin{center}
\caption{Data for EOS3 (SS2)}
\begin{tabular}{|c|c|c|c|c|c|c|}
\hline
$\rho_c~10^{14}$&$n_c/n_0$&$M/M_{\odot}$&R&$\omega_0~10^3$&$\omega_1~10^3$&$\omega_2~10^3$\\
g/c.c.&&&km&/sec.&/sec.&/sec.\\
\hline
17.17&6.067&0.423&5.070&86.879&195.416&299.090\\
18.17&6.382&0.539&5.460&74.016&172.966&264.894\\
19.17&6.695&0.659&5.794&63.013&153.699&236.761\\
20.17&7.006&0.781&6.078&53.079&136.745&212.463\\
21.17&7.298&0.855&6.227&47.332&127.643&199.063\\
22.17&7.588&0.923&6.351&42.074&119.663&186.932\\
23.17&7.876&0.986&6.453&37.131&112.402&176.070\\
24.17&8.156&1.036&6.524&33.069&106.694&167.790\\
25.17&8.428&1.075&6.575&29.646&102.152&161.274\\
26.17&8.699&1.110&6.615&26.321&98.030&155.270\\
27.17&8.967&1.142&6.647&22.992&94.193&149.689\\
28.17&9.227&1.167&6.667&20.163&91.206&145.298\\
29.17&9.485&1.188&6.682&17.335&88.518&141.539\\
30.17&9.741&1.207&6.693&14.307&86.006&137.932\\
31.17&9.994&1.224&6.691&10.827&83.642&134.554\\
32.17&10.242&1.237&6.702&6.982&81.741&131.825\\
33.17&10.488&1.249&6.703&$3.351$&79.959&129.293\\
35.17&10.974&1.270&6.698&$-$&76.694&124.666\\
40.17&12.148&1.301&6.650&$-$&70.680&116.215\\
45.17&13.278&1.316&6.622&$-$&66.380&110.292\\
50.17&14.371&1.323&6.573&$-$&63.141&105.918\\
55.17&15.537&1.325&6.518&$-$&60.616&102.546\\
\hline
\end{tabular}
\label{eos3}
\end{center}
\end{table*}

\begin{table*}[h]
\begin{center}
\caption{Data for bag model with B=60 \& ms=150}
\begin{tabular}{|c|c|c|c|c|c|c|}
\hline
$\rho_c~10^{14}$&$n_c/n_0$&$M/M_{\odot}$&R&$\omega_0~10^3$&$\omega_1~10^3$&$\omega_2~10^3$\\
g/c.c.&&&km&/sec.&/sec.&/sec.\\
\hline
6.20&2.421&0.691&8.549&38.964&92.549&142.426\\
7.20&2.778&1.019&9.544&27.409&73.557&114.644\\
8.20&3.122&1.240&10.021&20.733&63.870&100.636\\
9.20&3.454&1.393&10.263&15.915&57.854&92.024\\
10.20&3.776&1.501&10.393&11.910&53.693&86.161\\
11.20&4.089&1.581&10.452&7.900&50.628&81.883\\
12.20&4.396&1.639&10.469&2.159&48.272&78.626\\
13.20&4.695&1.683&10.462&$6.156$&46.395&76.048\\
15.20&5.277&1.741&10.405&$-$&43.526&72.203\\
17.20&5.839&1.775&10.321&$-$&41.429&69.514\\
23.70&7.560&1.805&10.012&$-$&37.316&64.404\\
\hline
\end{tabular}
\label{bag1}
\end{center}
\end{table*}

\begin{table*}[h]
\begin{center}
\caption{Data for bag model with B=75 \& ms=150}
\begin{tabular}{|c|c|c|c|c|c|c|}
\hline
$\rho_c~10^{14}$&$n_c/n_0$&$M/M_{\odot}$&R&$\omega_0~10^3$&$\omega_1~10^3$&$\omega_2~10^3$\\
g/c.c.&&&km&/sec.&/sec.&/sec.\\
\hline
9.83&3.573&1.072&8.923&24.809&73.494&115.482\\
10.83&3.892&1.198&9.148&20.030&67.175&106.403\\
11.83&4.203&1.293&9.281&16.097&62.623&99.944\\
12.83&4.506&1.366&9.356&12.565&59.194&95.089\\
13.83&4.804&1.422&9.396&9.063&56.480&91.314\\
14.83&5.095&1.467&9.412&4.502&54.280&88.272\\
15.83&5.318&1.502&9.411&$4.887$&52.470&85.769\\
20.83&6.748&1.594&9.294&$-$&46.505&77.834\\
25.83&8.031&1.622&9.123&$-$&43.109&73.610\\
28.83&8.769&1.626&9.020&$-$&41.638&71.868\\
\hline
\end{tabular}
\label{bag2}
\end{center}
\end{table*}

\end{document}